\documentstyle[twocolumn,prl,aps]{revtex}
\begin{document}

\title{Turning Bosons into Fermions:\\ Exclusion Statistics, Fractional
Statistics \\ and the Simple Harmonic Oscillator
}
\author{A.~D.~Speliotopoulos\cite{ads}}
\address{Institute of Physics, Academia Sinica, Nankang, Taipei, Taiwan}
\date{October 10, 1995}
\maketitle

\begin{abstract}
Motivated by Haldane's exclusion statistics, we construct creation
and annihilation operators for $g$-ons using a bosonic algebra. We
find that $g$-ons appear due to the breaking of a descrete symmetry of the
original bosonic system. This symmetry is intimately related to the
braid group and we demonstrate a link between
exclusion statistics and fractional statistics.

\end{abstract}


In 1991 Haldane proposed that excitations which obey fractional statistics
may exist in certain condensed matter systems even in
dimensions other than two $\cite{Hal}$. This was based on his
observation such excitations are present in the Calogero-Sutherland
model $\cite{Ha}$-$\cite{MiPoly}$. For these excitations, which were
dubbed ``$g$-ons'' by Nayak and Wilczek $\cite{nayak}$, he proposed
that the number of available one-particle states changes with the
increasing occupation of the state. Namely, if $d$ as the number of
one-particle states available for the particle and $N$ as its
occupation number, then
\begin{equation}
\Delta d = - g \Delta N
\end{equation}
where $g$ determines the statistics of the particle. Since the number
one-particle state for bosons does not change with $N$, $g=0$ for
bosons. For fermions, on the other hand, the number of available
one-particle states decreases by one with the addition of even one
fermion, giving a $g=1$. This definition of statistics has the
advantage of making no specific reference to the dimensionality of the
system which is being studied while the usual notion of fractional
statistics is often made in reference to anyons $\cite{anyons}$ which
traditionally have only been present in two spatial dimensions.

Since Haldane's work there have been a number of papers advancing his
ideas $\cite{nayak}$, $\cite{Wu1}$ - $\cite{FuKa}$ by using thermodynamic
arguments. This was done primarily by using the second of Haldane's
proposals: that the dimensionality of the Hilbert space of
$N$ $g$-ons is
\begin{equation}
D = \frac{[d+(1-g)(N-1)]!}{N![d-1-g(N-1)]!}.
\label{e0}
\end{equation}
This was used to construct a partition function for $g$-ons,
which was then compared with the partition function of various known
systems. Using this argument Murthy and Shanker was able to show that
anyons also obey exclusion statistics.

In another approach, Karabali and Nair have recently attempted to use
operator methods $\cite{Kara}$ to realize exclusion statistics
algebraicly. Specifically, they proposed the existence of creation and
annihilation operators $\tilde a$ and $\tilde a^\dagger$ for $g$-ons by
requiring $a^{m+1}=0$. The resultant Hilbert space is therefore
finite dimensional containing at most $m$ possible occupation
states. As the system looses a one-particle Hilbert space as
soon as the occupation number for the state changes by $m$, $g=1/m$.
Although $m$ ($1/g$) is necessarily an integer, since Nayak and
Wilczek have shown that the thermodynamical distribution functions
for $g$ and $1/g$ are related by a duality property, this limitation
does not seem to be to troublesome. Of greater concern is that since
a specific representation of these operators were not known, the
resulting commutation relations for $\tilde a$ and $\tilde a^\dagger$
could not be determined uniquely.

In this letter we shall extend Karabali and Nair's analysis by
constructing explicitly the creation and annihilation operators for
g-ons using a bosonic algebra. This is done by using the bosonic
number operator to define a operator whose eigenvalues are
abelian representations of the braid group and are commonly identified as
the ``statistics phases'' for anyons $\cite{anyons}$.
Projection operators are then constructed which projects states of
the bosonic Hilbert space $\cal H$ into states with definite
statistics phase. These projection operators are then used to define
the creation and annihilation operators. We shall further see from
this construction that $g$-ons appear in the bosonic system as the
result of the breaking of a descrete symmetry of the original bosonic
Hilbert space. In the spirit of Haldane's original work, this
analysis is done {\it without\/} reference to any specific spacetime
dimension.

Denoting the usual bosonic operators by $a$ and $a^\dagger$ and their
number operator by $N=a^\dagger a$, we begin with the unitary
operator
\begin{equation}
B_m \equiv \exp\left(\frac{2\pi i}{m+1}N\right),
\end{equation}
where $m$ is a non-negative integer. One can easily show that
\begin{equation}
B_m a B_m^\dagger = e^{-2\pi i/(m+1)} a\>,\quad {\cal
B}_m a^\dagger
B_m^\dagger = e^{2\pi i/(m+1)} a^\dagger.
\label{e2}
\end{equation}
Consequently, $[a,B_0]=0=[a^\dagger,B_0]$, and $B_0$ commutes with
every operator in the algebra generated by $a$
and $a^\dagger$. It is therefore (a Casimer operator) proportional to
the identity operator $I$. Since the eigenvalues of $N$ are the
non-negative integers, this proportionality constant is simply unity
$\cite{com}$.

Consider now $B_m$ for $m>0$ which has eigenvalues
\begin{equation}
e^{2\pi i j/(m+1)},
\label{e3}
\end{equation}
that are the $m+1$ roots of unity. Since only the ratio $j/(m+1)$
matter, it is understood from now on that $0\le j\le m$. It is also
known that for a fixed value of $j/(m+1)$, eq.~$(\ref{e3})$ is a one
dimensional (abelian) representation of the braid group
$e^{-i\pi\nu}$ if one identifies $\nu = -2j/(m+1)$ {}
$\cite{anyons}$. In this context, eq.~$(\ref{e3})$ is often also called
the ``statistics phase'' of an anyon. Consequently, we shall $B_m$
the quantum {\it braid \/} operator.

Next, $B_m \vert j + (m+1)q\rangle = e^{2\pi i j/(m+1)}\vert j +
(m+1)q\rangle$ where $\vert n\rangle$ is a state in $\cal H$ and
$q\ge0$ is an arbitrary integer. The bosonic Hilbert space $\cal H$
is also spanned by eigenstates of $B_m$. These states can be
isolated by using the projection operators
\begin{equation}
P^m_j= \frac{1}{m+1}\sum_{k=0}^m \exp\left(\frac{2\pi ik}{m+1}(N-j)\right),
\end{equation}
which have the properties
\begin{equation}
{P^m_j}^\dagger = P^m_j\>,\quad
P^m_j P^m_k = P^m_j \delta_{j,k}\>,\quad
\sum_{j=0}^m P^m_j = I,
\label{e5}
\end{equation}
and $P^m_{j\pm (m+1)} = P^m_{j}$. It is easily seen that $P^m_j
\vert n + (m+1)q\rangle =  \delta_{j,n}\vert n + (m+1)q\rangle$ and
the original Hilbert space decomposes into ${\cal H} ={\cal H}_0\oplus
\cdots \oplus {\cal H}_m$. Each state in ${\cal H}_j$ is an
eigenstate of $B_m$ with eigenvalue eq.~$(\ref{e3})$ and has
a definite statistics phase.

We then use $P^m_j$ to form the composite operators
\begin{equation}
e^m_j \equiv a P^m_j \>, \quad {e^m_j}^\dagger \equiv P^m_j
a^\dagger,
\label{e6}
\end{equation}
where $0\le j\le m$. Using eq.~$(\ref{e2})$,
\begin{equation}
P^m_j a = a P^m_{j+1}\>,\quad P^m_j a^\dagger = a^\dagger
P^m_{j-1},
\label{e6a}
\end{equation}
and we find
\begin{eqnarray}
e^m_j e^m_k = e^m_j e^m_{j+1} \delta_{k, j+1}\> &,&\quad
{e^m_k}^\dagger {e^m_j}^\dagger = {e^m_{j+1}}^\dagger {e^m_j}^\dagger
\delta_{k, j+1},
\nonumber \\
\{e^m_j, {e^m_k}^\dagger\} = T^m_j \delta_{j,k}\> &,&\quad
[T^m_j, T^m_k] = 0
\label{e7}
\end{eqnarray}
while
\begin{equation}
[e^m_j, T^m_{j+1}] = e^m_j + e^m_jN^e_j\>,\quad
[e^m_j, T^m_{j-1}] = -N^e_{j-1}e^m_j,
\end{equation}
and
$[e^m_j, T^m_k] =0$ for $\vert j-k\vert \ne1$. $N^e_j \equiv
{e^m_j}^\dagger e^m_j$, $N=\sum_j N^e_j$ and $[N^e_j,T^m_k] = 0$.
Written in this way the original background bosonic operators do not
explicitly appear, the algebra closes and no other
operators need to be introduced. These $e^m_j$ will be used to
construct the creation and annihilation operators for $g$-ons and
their resultant Hilbert spaces. To demonstrate that the
commutation relations eq.~$(\ref{e7})$ are sufficient to determine
the $g$-on Hilbert space upto an overall constant, we shall make no
further reference to the underlying bosonic algebra.

We begin with $m=1$, and $a^1_0 \equiv e^1_1$. Then from
eq.~$(\ref{e7})$, $(a^1_0)^2 = 0$ and
$\{a^1_0,{a^1_0}^\dagger\}=T^1_1$. Since, however,
$[a^1_0, T^1_1]=0$, if we restrict ourselves to only those operators
and Hilbert space generated by $a^1_0$ and ${a^1_0}^\dagger$ then
$T^1_1$ is once
again a multiple of the identity which can be set to unity by
re-scaling $a^1_0$. $a^1_0$ and ${a^1_0}^\dagger$ therefore obey the
usual {\it fermionic anticommutation\/} relations and generate a
fermionic Hilbert space.

Even though we still have the operator $e^1_0$, no other
interesting operator can be constructed for $m=1$ aside for the
trivial replacement of $e^1_0\leftrightarrow e^1_1$. Once
$e^1_0$ is also included in the sub-algebra the complete bosonic
Hilbert space will be reconstructed as can be seen from the
completeness relation in eq.~$(\ref{e5})$. We will no longer be
projecting $H$ into a sub-Hilbert space with a definite statistics
phase. This is a reflection of the very will known result that for
$m=1$ the statistics phase is $\pm1$ and only fermions or bosons can
be present.

For $m=2$ we construct the creation and annihilation operators
for $1/2$-ons by taking, for and arbitrary but fixed $j$,
\begin{equation}
a^2_j = e^2_{j+1} + e^2_{j+2}.
\end{equation}
{}$(a^2_j)^3=0$, as can easily be seen from the commutation
relations. To construct the Hilbert space, we take $\vert
\Omega\rangle^2_j$ as the ground state. It is also chosen
as an eigenstate of both $T^2_{j+1}$ and $T^2_{j+2}$ with eigenvalues
$\lambda^2_{j+1}$ and $\lambda^2_{j+2}$, respectively. This is
possible because these both operators commute among themselves as
well as the $1/2$-on number operator $N^2_j\equiv {a^2_j}^\dagger
a^2_j$. Then, because $e^2_{j+2}e^2_{j+1}=0$, one can show that
$\lambda^2_{j+2}=0$. The $1/2$-on Hilbert space is then spanned by
the normalized states
\begin{equation}
\vert \Omega\rangle^2_j\>,\quad
\frac{
	{a^2_j}^\dagger
     }{
     \sqrt{
     	\lambda^2_{j+1}
	  }
     }\vert
\Omega\rangle^2_j\>,\quad
\frac{
	({a_j^2}^\dagger)^2\vert \Omega\rangle^2_j
     }{
     \sqrt{
     	\lambda^2_{j+1}(\lambda^2_{j+1}+1)
	}
     }.
\end{equation}
They are eigenstates of $N^2_j = {a_j^2}^\dagger a^2_j$ with
eigenvalues $0$, $\lambda^2_{j+1}$ and $\lambda^2_{j+1}+1$,
respectively, and of $T^2_{j+1}$ with eigenvalues $\lambda^2_{j+1}$,
$\lambda^2_{j+1}$, and $0$. The constant $\lambda^2_{j+1}>0$ itself
cannot be determined by eq.~$(\ref{e7})$ alone. However, using the
underlying bosonic Hilbert space we find that $\lambda^2_{j+1} = j+1+3q$.

The generalization to arbitrary $m$ is straightforward. We take
\begin{equation}
a^m_j = \sum_{k=1}^m e^m_{j+k}\>,
\label{e10}
\end{equation}
for which $(a_j^m)^{m+1}=0$. $a^m_j$ and ${a^m_j}^\dagger$ are thus
creation and annihilation operators for $1/m$-ons. The $1/m$-on
Hilbert space is $m$-dimensional and spanned by the states
\begin{equation}
\vert p\rangle^m_j = \frac{
			({a^m_j}^\dagger)^p\vert\Omega\rangle^m_j
			}{
			  \sqrt{
			  	\lambda^m_{j+1}(\lambda^m_{j+1}+1)
				\cdots(\lambda^m_{j+1}+p-1)
			       }
			 },
\end{equation}
for $1\le p\le m$ with the ground state being
$\vert\Omega\rangle^m_j$. They are eigenstates of the $1/m$-on number
operator $N^m_j$ with eigenvalues $\lambda^m_{j+1}+p$, as well as of
$T^m_{j+1}\vert p\rangle^m_j = \lambda^m_{j+1}\delta_{p,1}\vert
p\rangle^m_j$. Once again $\lambda^m_{j+1}>0$ is an arbitrary
constant which, because $T^m_{j+1}\vert\Omega\rangle^m_j=
\lambda^m_{j+1}\vert\Omega\rangle^m_j$, ultimately depends on the
ground state of the system. For the underlying bosonic Hilbert space,
$\lambda^m_{j+1}=j+1+(m+1)q$.

In the $m\to\infty$ limit, eq.~$(\ref{e10})$ becomes an infinite sum,
and we find that $(a^\infty_j)^l\ne0$ for any finite $l$.
Using standard arguments we can show that
$\lambda^\infty_{j+1}=1$ and we simply recover the usual bosonic
algebra and Hilbert space. This can also be seen heuristically from
eq.~$(\ref{e3})$, by looking at the spectrum of $B_m$ for finite $m$
and taking the $m\to\infty$. We therefore identify $B_\infty=I$ and
note that $B_0=B_\infty$.

$B_m \vert p \rangle^m = e^{2\pi i p/(m+1)}\vert p \rangle^m$ for a
$B_m$ invariant ground state. Each $1/m$-on occupation state is
an eigenstate of the braid operator with the state containing $p$ of
the $1/m$-ons having a statistics phase $e^{2\pi ip/(m+1)}$. Each
$1/m$-on thereby has a statistics phase of $e^{2\pi i/(m+1)}$. For
$m=1$ this is just $-1$, as expected for fermions, while for
$m\to\infty$ it is $1$, as expected  for bosons.

We can understand the physics behind this construction of
$g$-ons by using the following symmetry arguments. We
first introduce the notion of $B_m$-Parity, which like the usual
parity $\cal P$ is a descrete symmetry and is generated by $B_m$.
Since, however, $B_m^{m+1}=I$, the eigenvalues of $B_m$ are in general
complex while $\cal P$ is a ${\bf Z}_2$ symmetry. The original
bosonic system is, of course, $B_m$-Parity invariant and $\cal H$
itself is spanned by states with definite $B_m$-Parity. In the
construction of the $g$-on operators $a^m_j$, ${a^m_j}^\dagger$,
however, only the projection operators $P^m_{j+1}, \dots, P^m_{j+m}$
were used. $P^m_j$ itself was purposely leftout. In effect, $a^m_j$
represents the the projection of any bosonic state into a
$m$-dimensional subspace spanned by these projection operators after which
$a$ is applied. $B_M$-Parity is explicitly broken by hand and
the states $\vert j+(m+1)q\rangle$ that $P^m_j$ projects into forms
the ground state $\vert\Omega\rangle^m_j$ for the $g$-on Hilbert
space. The choice of ground state for the g-ons is not unique and
there are an infinite number of ground states which can be used to
generate the Hilbert space, each corresponding to a different
$\lambda^m_{j+1}$. These ground states are {\it not\/}
equivalent, however, since the eigenvalues of $N^m_j$ are
$\lambda^m_{j+1}$ dependent.

Since the eigenvalues of $B_m$ are in general complex, $B_m$ is not
necessarily a physical observable. Its effect on the bosonic Hilbert
space is nevertheless dramatic and observable. The breaking of this
descrete symmetry reduces the infinite dimensional bosonic Hilbert
space into the finite dimensional $g$-on Hilbert space. Indeed, the
presence of $g$-ons in the system occurs precisely because this
descrete is broken in the original bosonic system.

As an explicit example of this, consider the case $m=1$, for which
$B_1 = (-1)^N$, $(B_1)^2=I$ and is a ${\bf Z}_2$ symmetry. With respect
to this operator, $\cal H$ consists of both even ($\vert
2n\rangle$) and odd ($\vert 2n+1 \rangle$) states. The fermionic operators
$a^1_0$ are, however,
constructed from $P^1_1$ only. $P^1_0$ was not, and could
not, be used or else the original Hilbert space would be reproduced.
$B_1$-Parity is explicitly broken and we find that the one fermion state
$(a^1_0)^\dagger\vert\Omega\rangle^1_0$ is odd under $B_1$. It has
statistics phase of $-1$, as expected. Also, if we identify $\cal
H$ with the one dimensional simple harmonic oscillator, then $B_1^1$
also functions as the usual parity operator. Equivalently, fermions
appear in the bosonic system due to parity being broken. This
breaking of parity is a well known effect for anyons. Unfortunately,
due to its complex eigenvalues $B_m$ does not have such a nice
physical interpretation for $m>1$.

To conclude, we have constructed the creation and annihilation
operators for g-ons; particles which obey Haldane's exclusion
statistics. This was done using the usual bosonic creation and
annihilation operators without any reference to a specific spacetime
dimension. Physically, $g$-ons appear in the bosonic system as the
result of the breaking of a descrete symmetry. Moreover, as the
construction explicitly used the braid operator $B_m$ whose
eigenvalues consist of abelian representations
of braid group, we have established a link between
Haldane's exclusion statistics, fractional statistics, the braid
group and anyons. Indeed, we have found that $g$-ons have a
statistics phase of $e^{2\pi ig/(g+1)}$, and have finite dimensional
Hilbert spaces, precisely as one would expect for anyons.
Consequently, denoting the usual statistics phase of an anyon by
$e^{\pi i \alpha}$, we see that $\alpha(g)=2g/(g+1)$, which has the
correct limiting values at $g=0,1$ for bosons and fermions,
respectively. (This this relation between $\alpha$ and $g$ is unique
only up to an even integer.) Notice also that $\alpha$ satisfies a
duality condition
\begin{equation}
\alpha(g)+\alpha(1/g)= 2,
\end{equation}
that is very similar to the partition function for $g$-ons obtained
through thermodynamical arguments.

Murthy and Shanker $\cite{Mur}$ has also shown that anyons obey
exclusion statistics. In their analysis a partition function for
$g$-ons was constructed using eq. $(\ref{e0})$ which was generalized to
infinite dimensional Hilbert spaces. A virial expansion is then
perform on this partition function and $g$ is shown to be very simply
related to the second virial coefficient in the high temperature
limit. Since the second virial coefficient has been calculated for
the anyon gas $\cite{virial1}$, they find that
$g_{ms}=\alpha_{ms}(2-\alpha_{ms})$. This is different from our
result of $g = \alpha/(2-\alpha)$. Their result was obtained via a
virial expansion, however, and, as they have pointed out, is valid for
a general anyon gas only if {\it all\/} the virial coefficients are
finite for the anyon gas, a result which is not yet known. Our results
would seem to suggest that either these virial coefficients
are not finite, or else the relationship they derived is valid only
in the high temperature limit near $\alpha_{ms}=0, 1$ when $\alpha
\approx\alpha_{ms}$. Notice also that $\alpha_{ms}$ does not satisfy
a simple duality relation between $g$ and $1/g$.

Traditionally, anyons have been associated with two dimensions where
the homotopy class $\pi_1(M_n)$ on the configuration space $M_n$ of $n$
hardcore particles is non-trivial. The intertwinning worldlines of
these particles in the Feynmann path integral formalism form
a representation of the abelian braid group. By using operator instead
of path integral methods to realize the braid group we have extended
the notion of anyons to arbitrary dimensions. There is, however, a
fundamental difference in the two approaches. In our approach
$g$-ons appear because an underlying descrete symmetry of
the bosonic Hilbert space is {\it broken}, while in the standard
description anyons are present precisely because the braid group is a
fundamental symmetry group which is {\it not\/} broken. This may also
be the cause of the differences between our $\alpha(g)$ and
$\alpha_{ms}(g)$. It would also be interesting to see if this symmetry
breaking can occur dynamically instead of by hand as we have done.

\newpage


\begin{references}
\bibitem[12]{com} This is a special case of a more general result. If
		  $f(x)$ is a periodic, analytic function, then
		  $[a,f(N)]=0$ and $f(N)=f(0) I$.
\end{references}
\end{document}